\documentclass[11pt]{article}
\usepackage{amssymb,latexsym,amsmath}
\usepackage{amsfonts, graphics}
\newtheorem{theorem}{Theorem}
\newtheorem{lemma}{Lemma}

\begin{document}
\title{Sharp bounds on the critical stability radius for relativistic charged spheres}
\author{H{\aa}kan Andr\'{e}asson\\
        Mathematical Sciences\\
        Chalmers and G\"{o}teborg University\\
        S-41296 G\"oteborg, Sweden\\
        email\textup{: \texttt{hand@math.chalmers.se}}}

\maketitle

\begin{center}
\textit{This work is dedicated to the memory of\\ my father Dan Andr\'{e}asson} (1933-2008).
\end{center}

\begin{abstract}
In a recent paper by Giuliani and Rothman \cite{GR}, the problem
of finding a lower bound on the radius $R$ of a charged sphere
with mass $M$ and charge $Q<M$ is addressed. Such a bound is
referred to as the critical stability radius. Equivalently, it can
be formulated as the problem of finding an upper bound on $M$ for
given radius and charge. This problem has resulted in a number of
papers in recent years but neither a transparent nor a general
inequality similar to the case without charge, i.e., $M\leq 4R/9,$
has been found. In this paper we derive the surprisingly
transparent inequality
\begin{equation}\nonumber
\sqrt{M}\leq\frac{\sqrt{R}}{3}+\sqrt{\frac{R}{9}+\frac{Q^2}{3R}}.
\end{equation}
The inequality is shown to hold for any solution which satisfies
$p+2p_T\leq\rho,$ where $p\geq 0$ and $p_T$ are
the radial- and tangential pressures respectively and $\rho\geq 0$
is the energy density. In addition we show
that the inequality is sharp, in particular we show that sharpness
is attained by infinitely thin shell solutions.
\end{abstract}

\section{Introduction}
Black holes for which the charge or angular momentum parameter equals the mass
are called extremal black holes. They are very central in black hole
thermodynamics due to their vanishing surface gravity and they represent
the absolute zero state of black hole physics. It is quite generally
believed that extremal black holes are disallowed by nature but a
proof is missing. One possibility to obtain an extremal black hole is to
produce one from the collapse of an already extremal object. Previous mainly
numerical studies
(\cite{dFSY2}, \cite{ARo}) have concluded that when $Q<M$ collapse always
takes place at a critical radius $R_c$ outside the outer horizon, and as
$Q$ approaches $M,$ this value approaches the horizon. This is similar to
the non-charged case where the Buchdahl inequality implies that collapse
will take place when $R<9M/4,$ i.e., $R_c=9M/4,$ cf. \cite{Bu1}. In the charged case
the critical value is expected to be smaller due to the Coulomb repulsion, 
and this is also shown to be the case below and in particular as $Q\to M$ 
the stability radius does approach the outer horizon. 
For more information on the relation of this topic to extremal black holes and
black hole thermodynamics we refer to \cite{ARo}, \cite{GR}, \cite{FH}
and \cite{BH} and the references therein. 

The problem of finding a similar bound as the classical Buchdahl
bound for charged objects have resulted in several papers; some of
these are analytical, cf. \cite{GR}, \cite{MDH}, \cite{FH},
\cite{FST}, \cite{HM} and \cite{YS}, whereas others are numerical
or use a mix of numerical and analytical arguments, cf.
\cite{ARo}, \cite{dFSY2}, and \cite{G} to mention some of them. We
refer the reader to the sources for the details of these studies
but in none of them a transparent bound has been obtained (except
in very special cases), on the contrary they have been quite
involved and implicit. Moreover, most of these studies rely on the
assumptions made by Buchdal, i.e., the energy density is assumed
to be non-increasing and the pressure to be isotropic.

In this work we will show that
\begin{equation}\label{Dan}
\sqrt{m_g}\leq\frac{\sqrt{r}}{3}+\sqrt{\frac{r}{9}+\frac{q^2}{3r}},
\end{equation}
given that $q<r$ (which is a physically natural assumption, cf.
the discussion below), and that $p+2p_T\leq\rho,$ where $p\geq 0$
and $p_T$ are the radial- and tangential pressures respectively
and $\rho\geq 0$ is the energy density. Here we have used lower
case letters $m_g,q$ and $r$ to stress that the inequality holds
anywhere inside the object. We refer to the equations (\ref{j})
and (\ref{mg}) below for the exact definitions of these quantities. 
To the best of our knowledge this bound has not appeared in 
the literature before. 

In the non-charged case a general proof of the Buchdahl inequality
$2m/r\leq 8/9,$ in the case when $p+2p_T\leq\rho,$ was first given
in \cite{An1}. A completely different proof was then given by
Stalker and Karageorgis \cite{KS} where also several other situations were
considered, e.g. the isotropic case where $p=p_T.$ The advantage
of the method in \cite{KS} (which is related to the method by Bondi
\cite{Bo} which however is non-rigorous) 
compared to the method in \cite{An1} is
that it is shorter and that it is more flexible in
the sense that other assumptions than $p+2p_T\leq \rho$ can be
treated. On the other hand the result in \cite{KS} is weaker than
the result in \cite{An1} in the sense that the latter method
implies that the steady state that saturates the inequality is
unique, it is an infinitely thin shell. Indeed, in \cite{An1} it is
shown that given any steady state, the value of $2m/r$ for this
state is strictly less than the value $2m/r$ of a state for which
the matter has been slightly re-distributed and this monotonic
property continues until an infinitely thin shell has been reached
for which $2m/r=8/9$. The method in \cite{KS} also shows sharpness
but only in the sense that there are steady states with $2m/r$
arbitrary close to $8/9,$ leaving open the possibility that
different kinds of steady states might share this feature.
Moreover, since the assumption $p+2p_T\leq \rho$ is satisfied by
solutions of the Einstein-Vlasov system it is natural to ask
if there exist regular static solutions to the \textit{coupled}
system which can have $2m/r$ arbitrary close to $8/9.$ This
question is given an affirmative answer in \cite{An2}, where in
particular it is shown that arbitrary thin shells which are
regular solutions of the spherically symmetric Einstein-Vlasov system
do exist. On the contrary, the matter quantities and the corresponding spacetimes
constructed in \cite{KS} for showing sharpness cannot be realized
by regular solutions of the Einstein-Vlasov system. The construction
in \cite{KS} gives that a solution which nearly saturates the inequality
$2m/r\leq 8/9$ satisfies $p+2p_T=\rho,$ and in addition
$p_T$ and $\rho$ are discontinuous. Neither of these two properties can be
realized by regular solutions of the (massive) Einstein-Vlasov system.

In the present work where we study charged objects we will 
adapt the method in \cite{KS} to show 
the inequality (\ref{Dan}) and its sharpness. This again supports the claim 
above that this method is very flexible. We have not been able to
carry out the strategy in \cite{An1} in this case. If we have
succeeded it would have given a more complete
characterization, cf. the discussion above. However, we do 
show in Theorem 2 below that an infinitely thin shell solution (with 
properties specified in the theorem) saturates 
the inequality, although we cannot show that no other steady states 
can saturate it as well. We also mention that in \cite{AE} a numerical 
study of the coupled Einstein-Maxwell-Vlasov system is carried out 
which supports that there are arbitrarily thin shell solutions for 
this system which saturate the inequality (\ref{Dan}). 

The outline of the paper is as follows. In the next section the
Einstein equations will be given and some basic quantities will be
introduced. In section 3 the main results are stated and section 4
is devoted to the proofs. In the final section we discuss our
inequality in view of the bound derived in \cite{GR} for a constant 
energy density profile. 
%and we briefly discuss a conjecture
%raised in \cite{KSA}. 

\section{The Einstein equations}
We follow closely the set up in \cite{GR} but here we also allow
the pressure to be anisotropic, i.e., the radial pressure $p$ and
the tangential pressure $p_T$ need not be equal. We assume throughout 
the paper that $p,$ the energy density
$\rho,$ and the charge density $j^0$ are non-negative. We study
spherically symmetric mass and charge distributions and we write
the metric in the form 
\begin{displaymath}
ds^{2}=-e^{2\mu(r)}dt^{2}+e^{2\lambda(r)}dr^{2}
+r^{2}(d\theta^{2}+\sin^{2}{\theta}d\varphi^{2}),
\end{displaymath}
where $r\geq 0,\,\theta\in [0,\pi],\,\varphi\in [0,2\pi].$
%Asymptotic flatness is expressed by the boundary conditions
%\begin{displaymath}
%\lim_{r\rightarrow\infty}\lambda(r)=\lim_{r\rightarrow\infty}\mu(r)
%=0,
%\end{displaymath}
%and a regular centre requires
%\begin{displaymath}
%\lambda(0)=0.
%\end{displaymath}
It is well-known that the Reissner-Nordstr\"{o}m solution for the
charged spherically symmetric case gives
\begin{equation}\label{boundary}
e^{-2\lambda(r)}=1-\frac{2M}{r}+\frac{Q^2}{r^2}=e^{2\mu(r)},\;
r\geq R.
\end{equation}
Here $R$ is the outer radius of the sphere and $Q$ is the total
charge. This solution is a vacuum solution. The purpose of this
work is to investigate the behaviour of $\lambda$ and $\mu$ when
the matter and charge densities are non-zero for $r<R.$
Before writing down
the Einstein equations let us introduce some quantities following
\cite{GR}. Let
\begin{equation}\label{j}
q(r)=4\pi \int_0^r e^{(\lambda+\mu)(\eta)}\,\eta^2 j^0\,d\eta,
\end{equation}
and
\begin{equation}\label{mi}
m_i(r)=4\pi \int_0^r \eta^2\rho\, d\eta,
\end{equation}
where $q(r)$ is the charge within the sphere with area radius $r$ and
$m_i(r)$ is the mass within this sphere. The subscript $i$ is
used to distinguish $m_i$ from the gravitational mass $m_g$ which
is defined below. Let us also introduce the quantity
\[ F(r)=\int_0^r \frac{q^2(\eta)}{\eta^2} d\eta. \]
The Einstein equations for $\lambda$ and $\mu$ now read (cf.
\cite{ARo} and \cite{GR})
\begin{equation}\label{ee1}
\frac{1}{r^2}+\frac{2\lambda_r
e^{-2\lambda}}{r}-\frac{e^{-2\lambda}}{r^2}=
8\pi\rho+\frac{q^2(r)}{r^4},
\end{equation}
and
\begin{equation}\label{ee2}
\frac{1}{r^2}-\frac{2\mu_r
e^{-2\lambda}}{r}-\frac{e^{-2\lambda}}{r^2}= -8\pi
p+\frac{q^2(r)}{r^4},
\end{equation}
where the subscript $r$ denotes differentiation with respect to
$r.$ Equation (\ref{ee1}) can be written as
\begin{equation}\label{ee1alt}
\frac{d(e^{-2\lambda}r)}{dr}=1-8\pi r^2\rho-\frac{q^2(r)}{r^2},
\end{equation}
so that
\begin{equation}\label{e-2l}
e^{-2\lambda}=1-\frac{2m_i(r)}{r}-\frac{F(r)}{r}.
\end{equation}
By requiring that (\ref{e-2l}) matches the exterior solution
(\ref{boundary}) at $r=R$ gives
\begin{equation}\nonumber
1-\frac{2M}{R}+\frac{Q^2}{R^2}=1-\frac{1}{R}\int_0^{R}(8\pi\rho
\eta^2 +\frac{q^2}{\eta^2})d\eta
\end{equation}
or
\[ M=\frac12\int_0^{R}(8\pi\rho \eta^2
+\frac{q^2}{\eta^2})d\eta+\frac{Q^2}{2R}, \] which defines the
total gravitational mass $M$. In view of this relation we now define
the gravitational mass $m_g$ within a given area radius $r$ by 
\begin{equation}\label{mg}
m_g(r)=m_i(r)+\frac{F(r)}{2}+\frac{q^2(r)}{2r}.
\end{equation}
In terms of the gravitational mass we thus get
\begin{equation}\label{e-2lg}
e^{-2\lambda(r)}=1-\frac{2m_g(r)}{r}+\frac{q^2(r)}{r^2}.
\end{equation}
Let us also write down the Tolman-Oppenheimer-Volkov equation
which follows from the Einstein equations, cf. \cite{ARo}, but
note that in our case $p$ is allowed to be different from $p_T$
which modifies the equation accordingly
\begin{equation}\label{TOV}
p_{r}=\frac{qq_r}{4\pi r^4}+\frac2r (p_T-p)-(\rho+p)e^{2\lambda}
\big(\frac{m_g(r)}{r^2}+4\pi rp-\frac{q^2}{r^3}\big).
\end{equation}

\section{Set up and main results}
The problem of finding an upper bound on the total gravitational mass
that a
sphere of area radius $R$ with total charge $Q$ can hold, or
equivalently, to find the smallest radius $R_c,$ referred to as
the critical stability radius, for which a physically acceptable
solution of the Einstein equations can be found, is formulated in
\cite{GR} as follows:

A physically acceptable solution should satisfy
\begin{eqnarray}
& &\rho\geq 0,\; p\geq 0\; \mbox{ and }\; \mu>-\infty,\label{inconditions1}\\
& &0\leq Q<M, \; R>R_+.\label{inconditions12}
\end{eqnarray}
Here $R_+=M+\sqrt{M^2-Q^2}$ is the outer horizon of a
Reissner-Nordstr\"{o}m black hole. The quantities $m_g$ and $q$
should satisfy
\begin{eqnarray}
& &m_g(R)=M,\; q(R)=Q,\label{inconditions2}\\
& &q\leq m_g, \; m_g+\sqrt{m_g^2-q^2}<r.\label{inconditions22}
\end{eqnarray}
We see immediately that these relations imply that $q/r\leq m_g/r<1.$ 
%Given a regular solution (defined below) we define which in particular implies 
%that $q(r)/r=0 at r=0$ and that 
%\begin{equation}\label{sq}
%s_q:=\inf_{r\geq 0}-\frac{q^2(r)}{r^2}, 
%\end{equation}
%and since the solutions that we are interested in satisfy 
%$q/r\to 0$ as $r\to\infty$ we must have 
%\begin{equation}\label{sqlessthan1}
%s_q>-1. 
%\end{equation}
We will in addition assume that the following condition holds
\begin{equation}\label{energyrel}
p+2p_T\leq\rho.
\end{equation}
The condition (\ref{energyrel}) is likely to be satisfied for most
realistic matter models, cf. \cite{Bo2}, and in particular it
holds for Vlasov matter, cf. \cite{An4} for more information on this matter model. 
\\
\textit{Remark. }In \cite{An1} and \cite{An3} the following generalization of
this condition was imposed, namely that
\begin{equation}\label{energyrel2}
p+2p_T\leq\Omega\rho, \mbox{ for some } \Omega\geq 0.
\end{equation}
However, in contrast to the non-charged case where a bound on $M$ is given by
a simple formula depending on $\Omega$ the simplicity is completely lost
in the charged cased except when $\Omega=1.$ Now, the case $\Omega=1$ should 
be considered as 
the principal case, cf. \cite{Bo2}, and in the non-charged case it is when
$\Omega=1$ that the classical bound $2m/r<8/9$ is recovered.

We are now almost ready to state our main result but first we define what
we mean by a regular solution of the spherically symmetric
Einstein equations. We say that $\Psi:=(\mu,\lambda,\rho,p,p_T,j^0)$ is a 
regular solution if the matter quantities $\rho,p,p_T$ and $j^0$ are bounded 
everywhere and $C^1$ except possibly at finitely many points, 
$p$ has compact support and the equations
(\ref{j}), (\ref{ee1}), (\ref{ee2}) and (\ref{TOV}) are satisfied
(where the matter quantities are $C^1$) and the constraints
(\ref{inconditions1}) and (\ref{inconditions22}) are satisfied.
\begin{theorem}\label{thm1}
Let $\Psi$ be a regular solution of the Einstein equations
and assume that (\ref{energyrel}) holds. Then
\begin{equation}\label{chineq}
\sqrt{m_g(r)}\leq \frac{\sqrt{r}}{3}+\sqrt{\frac{r}{9}+\frac{q^2(r)}{3r}}.
\end{equation}
Moreover, the inequality is sharp in the subclass of regular solutions for
which $p_T\geq 0.$
\end{theorem}

Let us immediately make a consistency check so that (\ref{chineq}) 
ensures that the stability radius is strictly outside the outer horizon. 
Thus we wish to show that the inequality 
(\ref{chineq}) implies that 
$e^{-2\lambda(r)}=1-\frac{2m_g}{r}+\frac{q^2}{r^2}>0,$ or equivalently that 
$$\sqrt{\frac{m_g}{r}}<\sqrt{\frac12+\frac{q^2}{2r^2}}.$$ 
In view of inequality (\ref{chineq}) this holds if 
$$\frac{1}{3}+\sqrt{\frac{1}{9}+\frac{q^2}{3r^2}}< 
\sqrt{\frac12+\frac{q^2}{2r^2}}.$$ An elementary computation shows that this 
is true as long as 
$$\frac{q^2}{r^2}< 1,$$ which always holds. 

The proof of Theorem \ref{thm1} relies on the method in \cite{KS}
for the non-charged case. In the introduction we discussed the
strength of this method but also its shortages; the question of
uniqueness of the steady state that saturates the inequality
(\ref{chineq}) is left open, and the constructed steady states
that nearly saturate the inequality cannot be a solutions of the
coupled Einstein-Vlasov system (these issues were answered in
\cite{An1} and \cite{An2} respectively). Furthermore, it is not
completely obvious from the construction in \cite{KS} that these
solutions approach an infinitely thin shell. This point also
carries over in our proof of sharpness in Theorem \ref{thm1} and
we therefore find it natural to include a proof of the fact that
an infinitely thin shell does saturate (\ref{chineq}), although we
have not been able to adapt the strategy in \cite{An1} to show
that no other steady state can have this property. Furthermore,
the numerical study in \cite{AE} supports that the maximizer for
the spherically symmetric Einstein-Vlasov-Maxwell system is an
infinitely thin shell.

We therefore investigate a sequence of regular shell solutions
which approach an infinitely thin shell and adapt the method in
\cite{An3}. More precisely, let $\Psi_k$ be a sequence of regular
solutions such that $p_k,(p_T)_k,j^0_k$ and $\rho_k$ 
have support in $[R_k,R].$ Denote by $M_k$ the total gravitational mass and by
$Q_k$ the total charge of the corresponding solution in the sequence 
and assume that $Q:=\lim_{k\to\infty}Q_k,$ and 
$M=\lim_{k\to\infty}M_k$ exist 
and that $\sup_{k}q_k/r<1$. Furthermore, assume that 
$\int_{R_k}^{R} r^2 p_kdr\to 0,$ and 
$\int_{R_k}^{R} r^2(2(p_T)_k-\rho_k)\, dr\to 0$ as $k\to\infty.$ 
\begin{theorem}\label{thmshells}
Assume that $\{\Psi_k\}_{k=1}^{\infty}$ is a sequence of regular
solutions with support in $[R_k,R]$ with the properties specified above 
and assume that 
\begin{equation}
\lim_{k\to\infty}\frac{R_k}{R}=1.\label{hypothesis}
\end{equation}
Then 
\begin{equation}
\sqrt{M}=
\frac{\sqrt{R}}{3}+\sqrt{\frac{R}{9}+\frac{Q^2}{3R}}.\label{chineq2}
\end{equation}
\end{theorem}
\textit{Remark. }That sequences exist with these properties, in
particular the property (\ref{hypothesis}), has been proved for 
the (non-charged) Einstein-Vlasov system, cf. \cite{An2} (and
\cite{AR2} for a numerical study). The investigation carried out
in \cite{AE} also supports that such sequences exist for 
the Einstein-Vlasov-Maxwell system. 

\section{Proofs}

\textbf{Proof of Theorem \ref{thm1}. }
As described above our method of proof is an adaption of the method in \cite{KS} to
the charged case. Let a regular solution be given and let us define
\begin{equation}\label{mlambda}
m_\lambda(r)=m_i(r)+\frac{F(r)}{2}=m_g-\frac{q^2}{2r},
\end{equation}
and let $$x\equiv \frac{2m_{\lambda}}{r},\;\; y\equiv 8\pi r^2p,\;\;
z\equiv \frac{q^2}{r^2}.$$ Note that the conditions (\ref{inconditions1}) 
and (\ref{inconditions22}) imply that
\begin{equation}\label{xyzboundsthankstoq}
x<1,\; y\geq 0,\mbox{ and } z<1. 
\end{equation}
Indeed, the two latter bounds are immediate and the former follows since
(\ref{inconditions22}) gives that $q^2>m_g^2-(r-m_g)^2$ so that
\begin{equation}\label{xboundbyGRass}
x=\frac{2m_{\lambda}}{r}=\frac{2m_g^2}{r}-\frac{q^2}{r^2}
<\frac{2m_g}{r}-\frac{m_g^2-(r-m_g)^2}{r^2}=1.
\end{equation}
\begin{lemma}\label{lemmacurve}
The variables $(x,y,z)$ give rise to a parametric curve in
$[0,1)\times [0,\infty)\times [0,1)$ and
satisfy the equations
\begin{eqnarray}
& &\displaystyle 8\pi r^2\rho=2\dot{x}+x-z,\\
& &\displaystyle 8\pi r^2p=y,\\
& &\displaystyle 8\pi
r^2p_T=\frac{x+y-z}{2(1-x)}\dot{x}+\dot{y}-\dot{z}-z+\frac{(x+y-z)^2}{4(1-x)},
\end{eqnarray}
where the dots denote derivatives with respect to
$\beta:=2\log{r}.$
\end{lemma}
\textit{Proof of Lemma \ref{lemmacurve}: }
The proof is a straightforward computation using the Einstein equations (\ref{ee1})
and (\ref{ee2}) and the Tolman-Oppenheimer-Volkov equation (\ref{TOV}).
\begin{flushright}
$\Box$
\end{flushright}

Now let $$w(x,y,z)=\frac{(3(1-x)+1+y-z)^2}{1-x}.$$
Differentiating with respect to $\beta$ gives
\begin{equation}\label{wdiff}
\dot{w}=\frac{4-3x+y-z}{(1-x)^2}\big[(3x-2+y-z)\dot{x}+2(1-x)\dot{y}+2(1-x)\dot{z}\big].
\end{equation}
Now, using the expressions of the matter terms given in Lemma
\ref{lemmacurve} the condition $p+2p_T\leq\rho$ can be written
\begin{equation}\label{doteq}
(3x-2+y-z)\dot{x}+2(1-x)(\dot{y}-\dot{z})\leq\frac{-\alpha(x,y,z)}{2},
\end{equation}
where $$\alpha=3x^2-2x+(y-z)^2+2(y-z).$$ From (\ref{wdiff}) and (\ref{doteq}) it now
follows that
\begin{eqnarray}
\dot{w}&=&\frac{4-3x+y-z}{(1-x)^2}\big[(3x-2+y-z)\dot{x}+2(1-x)\dot{y}+2(1-x)\dot{z}\big]\nonumber\\
&\leq& -\frac{4-3x+y-z}{2(1-x)^2}\;\alpha(x,y,z).
\end{eqnarray}
Since $0\leq x<1,\;\; y\geq 0$ and $0\leq z<1,$ it follows that
$w$ is decreasing whenever $\alpha>0,$ which implies that
\begin{equation}
w\leq\max_E w(x,y,z),
\end{equation}
where $$E=\{(x,y,z):0\leq x\leq 1,\,y\geq 0,\,0\leq z\leq 1\mbox{
and }\alpha(x,y,z)\leq 0\}.$$ To solve this optimization problem
we introduce $s=y-z$ and note that $\max_E w(x,y,z)=\max_{E'} w(x,s)$ where
$$E'=\{x,s):0\leq x\leq 1,\, s\geq -1\mbox{ and }\alpha(x,s)\leq
0\}.$$ It is straightforward to conclude that there are no
stationary points in the interior of $E',$ so the maximum is
attained at the boundary $\partial E'$ of $E'.$ The Lagrange
multiplier method leads to the following system of equations
\begin{eqnarray}
& &(1-x)(6(1+s)+4(3x-1))+2(1+s)^2=0,\label{ekv1}\\
& &x(3x-2)+s(s+2)=0.\label{ekv2}
\end{eqnarray}
From (\ref{ekv2}) we have that $s^2=-2s-x(3x-2)$ which substituted
into (\ref{ekv1}) results in the equation
\begin{equation}\label{factor}
(x+s)(1-x)=0.
\end{equation}
If $x=-s$ we get from (\ref{ekv2}) that $4s^2+4s=0$ so that either
$s=0=x$ or $s=-1=-x.$ In the latter case we get $w(1,-1)=0,$ and
the former case gives $w(0,0)=16.$ We thus conclude that $w\leq
16$ throughout the curve. Since $p\geq 0$ it follows from the inequality
$w\leq 16$ that
\begin{equation}\label{pnoll}
\displaystyle
\big(3(1-\frac{2m_g}{r}+\frac{q^2}{r^2})+1-\frac{q^2}{r^2}\big)^2\leq
16\big(1-\frac{2m_g}{r}+\frac{q^2}{r^2}\big).
\end{equation}
This is easily seen to be equivalent to
\begin{equation}\label{square}
\big(\frac{6m_g}{r}-\frac{2q^2}{r^2}\big)^2\leq\frac{16m_g}{r}.
\end{equation}
Taking the square root of both sides and rearranging leads to
\begin{equation}
\displaystyle\big(\sqrt{m_g}-\frac{\sqrt{r}}{3}-\sqrt{\frac{r}{9}+\frac{q^2}{3r}}\big)
\big(\sqrt{m_g}-\frac{\sqrt{r}}{3}+\sqrt{\frac{r}{9}+\frac{q^2}{3r}}\big)\leq
0.
\end{equation}
Since the second bracket is always non-negative and vanishes only
if $m_g=q=0$ we have
\begin{equation}\label{ineqQproof}
\sqrt{m_g}-\frac{\sqrt{r}}{3}-\sqrt{\frac{r}{9}+\frac{q^2}{3r}}\leq
0,
\end{equation}
which is the first claim.

To show sharpness
\begin{figure}[htbp]
\begin{center}
\scalebox{.5}{\includegraphics{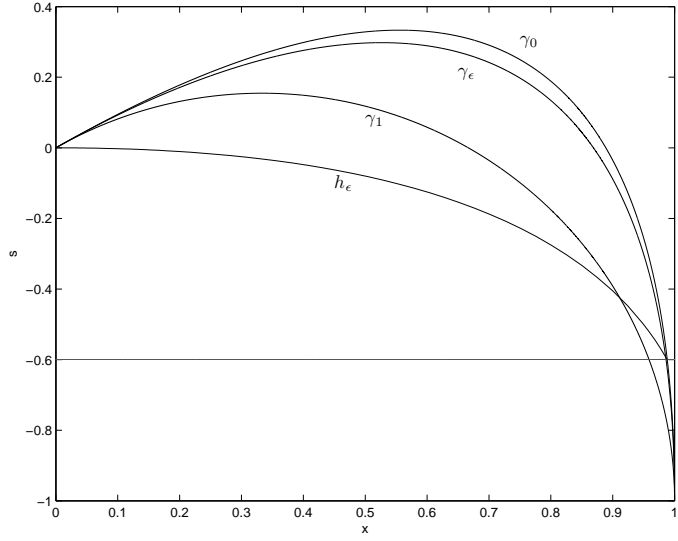}}
\end{center}
\caption{The curves $\gamma_0$, $\gamma_1,$ $\gamma_\epsilon$ and $h_{\epsilon}.$}
\end{figure}
we will construct a spacetime such that the corresponding curve
from Lemma \ref{lemmacurve} intersects a small neighbourhood of
$(x_q,0,z_q),$ where $z_q<1$ is a given ratio $q^2/r^2$ and $x_q$
is the corresponding value of $x$ when equality holds in
(\ref{ineqQproof}), i.e.,
$$x_q:=\frac{4}{9}-\frac{z_q}{3}+\frac43\sqrt{\frac19+\frac{z_q}{3}}.$$
We will construct such a spacetime by showing that there exists a
curve
$$x=x(\tau),\;\;y=y(\tau),\;\;z=z(\tau);\;\; \tau\in [0,\infty),$$
which passes near $(x_q,0,z_q)$ and in addition has the following
properties
\begin{itemize}
\item (A1) $\frac{1}{\alpha}\frac{dw}{d\tau}$ is negative and
locally integrable, \item (A2) x(0)=y(0)=z(0)=0, \item (A3) $0\leq
x(\tau)<x_q,\;z(\tau)\leq z_q,$ \item (A4) $y(\tau)=0$ for all
large enough $\tau$, $x(\tau)\to 0$ and $z(\tau)\to 0$ as
$\tau\to\infty,$ \item (A5) the curve is $C^1$ except for finitely
many points.
\end{itemize}
Below we will denote $s=y-z$ as above and the curves
$(x(\tau),y(\tau),z(\tau))\in [0,1)\times [0,\infty)\times [0,1),$
and $(x(\tau),y(\tau),s(\tau))\in [0,1)\times [0,\infty)\times
(-1,\infty)$ will be used interchangeably. Let us first see that
if we have a curve which satisfies (A1)-(A5) a spacetime can be
constructed. Indeed, let
\begin{equation}\label{kappa}
\kappa(\tau)=-\frac{1}{\alpha(x,s)}\frac{dw}{d\tau}\frac{2(1-x)^2}{4-3x+s},
\end{equation}
and observe that $\kappa$ is positive and locally integrable by
(A1) and (A2). Next define
\begin{equation}\label{beta}
\beta=\int\kappa\,d\tau,
\end{equation}
and
\begin{equation}\label{r}
r=e^{\beta/2}.
\end{equation}
and define the metric coefficients by
\begin{eqnarray}\label{lambdamu}
& &\lambda=-\frac{1}{2}\log{(1-x)}\\
& &\mu=-\int\frac{x+y}{4(1-x)}\kappa\,d\tau.
\end{eqnarray}
It is straightforward to check that $\lambda$ and $\mu$ solve the
Einstein equations (\ref{ee1}) and (\ref{ee2}). 
The definition of $\kappa$ now implies
\begin{equation}\label{wdot2}
\dot{w}=\frac{1}{\kappa}\frac{dw}{d\tau}=-\frac{4-3x+s}{2(1-x)^2}\alpha(x,s),
\end{equation}
where we recall that dots denote differentiation with respect to
$\beta=2\log{r}.$ Using (\ref{wdiff}) we thus have
\begin{equation}\label{mattersat}
(3x+s-2)\dot{x}+2(1-x)\dot{s}=-\frac{\alpha(x,s)}{2},
\end{equation}
which is equivalent to the relation $p+2p_T=\rho$ in view of
(\ref{doteq}). We will now show that such a curve exists. Let
us fix some small $\epsilon>0$ and define
\begin{equation}\label{wepsilon}
w_{\epsilon}(x,s):=\frac{((3-3\epsilon)(1-x)+1+s))^2}{1-x}.
\end{equation}
Consider now the curve $\gamma_{\epsilon}$ in the $(x,s)$-plane
defined by
\begin{equation}\label{epscurve}
w_{\epsilon}(x,s)=(\epsilon\sqrt{1+3x}+4(1-\epsilon))^2.
\end{equation}
Define the corresponding curve in $\mathbb{R}^3$ by 
\begin{equation}\label{xyz}
(x,y,z)=(x,\max{(0,s)},\max{(0,-s)}),
\end{equation}
so that $s=y-z.$ In figure 1 the curve $\gamma_{\epsilon}$ is
depicted together with the curves $\gamma_0$ and $\gamma_1.$ Note
that $\gamma_0$ is the curve $w(x,s)=16$ which passes through
$(0,0)$ and $(1,-1).$ The curve $\gamma_1$ is the curve $\alpha(x,s)=0$ 
which also passes through $(0,0)$ and $(1,-1).$ The dotted line shows 
the line $s=s_q:=-z_q$ (for the choice $z_q=0.6$) and the intersection of 
the curve $\gamma_0$ with this line is the point $(x_q,s_q)$. 
It is clear that for a sufficiently small $\epsilon>0$ the curve
$\gamma_{\epsilon}$ intersects an arbitrarily small neighbourhood
of $(x_q,s_q).$ Let us denote the point of intersection of
$\gamma_{\epsilon}$ and the line $s=s_q$ by 
$(x^{\epsilon}_q,s_q).$ Let us now define 
$\Gamma:=\gamma_{\epsilon}+h_{\epsilon},$ where $h_{\epsilon}$ is
the curve given by the equation
\begin{equation}\label{odexs}
\frac{ds}{dx}=\frac{2s}{s+x}, \mbox { such that }s(x^{\epsilon}_q)=s_q.
\end{equation}
It is clear from the defining equation that $h_{\epsilon}\in\{(x,s):x\geq 0, s\leq 0,
s+x>0\}$ and that the solutions approach the point $(0,0)$ for all
admissible starting points $(x^{\epsilon}_q,s_q)$ (note that 
$x_q+s_q>0$). The curve $h_\epsilon$ is depicted 
in figure 1. 
%shows that the solutions approach the point $(0,0)$ for all
%admissible starting points $(x^{\epsilon}_q,s_q)$. It is also 
%clear that it cannot intersect the x-axis since $ds/dx=0$ there. 

It remains to show that (A1)-(A5) are satisfied for the curve $\Gamma$ 
and that $\rho,p$ and $p_T$ are non-negative along the curve. 
It is obvious that $\Gamma$ satisfies (A2)-(A5). To see that 
it satisfies (A1) we first consider the first part of the curve $\gamma_{\epsilon}$ 
and note that 
$\alpha>0$ along $\gamma_{\epsilon}.$ This follows since
$\gamma_{\epsilon}$ lies above $\gamma_1$ and $\alpha=0$ along
$\gamma_1$ and 
$$\frac{\partial\alpha}{\partial s}=2s+2>0,\mbox{ for }s>-1.$$
Hence it is sufficient to show that $dw/d\tau<0$ to establish that
$$\frac{1}{\alpha}\frac{dw}{d\tau}<0.$$
We differentiate (\ref{epscurve}) and obtain
\begin{equation}\label{dwepsdtau}
\frac{dw_{\epsilon}}{d\tau}=\frac{3\epsilon}{\sqrt{1+3x}}\frac{3(1-\epsilon)(1-x)+1+s}{\sqrt{1-x}}\frac{dx}{d\tau}.
\end{equation}
If we now differentiate (\ref{wepsilon}) directly we get
\begin{equation}\label{wepsdiff}
\frac{dw_{\epsilon}}{d\tau}=\frac{(3-3\epsilon)(1-x)+1+s}{(1-x)^2}\big[(-(3-3\epsilon)(1-x)+1+s)\frac{dx}{d\tau}+2(1-x)\frac{ds}{d\tau}\big].
\end{equation}
Comparing (\ref{dwepsdtau}) and (\ref{wepsdiff}) gives
\begin{equation}\label{compare}
2(1-x)\frac{ds}{d\tau}=\big(\frac{3\epsilon
(1-x)^{3/2}}{\sqrt{1+3x}}+3(1-\epsilon)(1-x)-1-s\big)\frac{dx}{d\tau}.
\end{equation}
Thus differentiating $w$ along $\gamma_{\epsilon},$ substituting
for $ds/d\tau$ using (\ref{compare}), leads to
\begin{equation}\label{dwdtau}
\frac{dw}{d\tau}=\frac{3\epsilon(4-3x+s)}{1-x}\frac{\sqrt{1-x}-\sqrt{1+3x}}{\sqrt{1+3x}}\frac{dx}{d\tau}.
\end{equation}
Since $dx/d\tau>0$ along $\gamma_{\epsilon}$ and since $0\leq x<1$
and $s>-1$ we get that $$\frac{dw}{d\tau}<0.$$ It remains to show
that $\alpha^{-1}dw/d\tau$ is negative also along the curve
$h_{\epsilon}.$ Here we have
\begin{eqnarray}\label{dwdtauh}
\displaystyle \frac{1}{\alpha}\frac{dw}{d\tau}&=&\frac{3(1-x)+1+s}{(x(3x-2)+s(s+2))(1-x)^2}\big[(3x-2+s)+2(1-x)\frac{ds}{dx}\big]\frac{dx}{d\tau}\nonumber\\
\displaystyle&=&\frac{[3(1-x)+1+s]}{(x+s)(1-x)^2}\frac{dx}{d\tau},
\end{eqnarray}
where we used (\ref{odexs}) for $ds/dx.$ Since $dx/d\tau<0$ along
$h_{\epsilon}$ the claim follows since $x+s>0$ along $h_{\epsilon}$. 
Hence, 
\begin{equation}
\frac{1}{\alpha}\frac{dw}{\tau}<0, \end{equation} along $\Gamma.$
The local integrability of this expression follows by inspection
of the formulas above since $0\leq x\leq x^{\epsilon}_q<1$. Thus
condition (A1) holds along $\Gamma$. Finally we show that $\rho,p$
and $p_T$ are non-negative along $\Gamma.$ Since $y\geq 0$ along
$\Gamma,$ cf. (\ref{xyz}), it immediately follows that $p\geq 0$. 
Since (\ref{mattersat}) implies that $\rho=p+2p_T$ we only need 
to show that $p_T\geq 0$ along $\Gamma$. First we consider the first 
part of $\Gamma,$ i.e., the curve $\gamma_{\epsilon}.$ 
From Lemma \ref{lemmacurve} and (\ref{compare}) we have 
\begin{eqnarray}\label{pTgeqzero}
8\pi
r^2p_T&=&\frac{x+s}{2(1-x)}\dot{x}+\dot{s}-z+\frac{(x+s)^2}{4(1-x)}\nonumber\\
&=&\big(\frac{3\epsilon\sqrt{1-x}}{2\sqrt{1+3x}}+1-\frac32\epsilon\big)\frac{1}{\kappa}\frac{dx}{d\tau}+[-z+\frac{(x+s)^2}{4(1-x)}].
\end{eqnarray}
Since $\epsilon$ is small the first term is positive since
$dx/d\tau>0$ along $\gamma_{\epsilon}.$ The term in square
brackets can also be seen to be positive. Indeed, along the part of
$\gamma_{\epsilon}$ where $s\geq 0,$ $z=0$ and the claim is trivial so
we focus on the part where $s<0.$ Here $z=-s$ and we thus want
to show that
\begin{equation}\label{mzps}
s+\frac{(x+s)^2}{4(1-x)}> 0
\end{equation}
along the part of $\gamma_\epsilon$ where $s<0.$ We evaluate 
the left hand side of (\ref{mzps}) along $\gamma_0$ and show 
that it is positive there and then we conclude by continuity 
that this statement also holds along $\gamma_{\epsilon}$ for 
$\epsilon$ small. Along $\gamma_0$ we have the relation 
\begin{equation}\label{xrels}
x=\frac{4+3s}{9}+\frac{4}{3}\sqrt{\frac{1}{9}-\frac{s}{3}}.
\end{equation}
A straightforward calculation now gives that along 
$\gamma_0$ 
$$s+\frac{(x+s)^2}{4(1-x)}=s+4(\frac{1}{3}+\sqrt{\frac{1}{9}-\frac{s}{3}})^2\geq 
s+\frac{16}{9}>0.$$ Hence $p_T>0$ also along $\gamma_{\epsilon}$ for a sufficiently 
small $\epsilon$. Along $h_{\epsilon}$ it holds by construction, cf. (\ref{odexs}) 
and Lemma (\ref{lemmacurve}), that 
$\rho=0$ and thus $p=p_T=0.$ This completes the proof of Theorem \ref{thm1}. 

\begin{flushright}
$\Box$
\end{flushright}

\textbf{Proof of Theorem \ref{thmshells}. }Let us begin with a few
general facts. Consider a regular solution where the matter quantities are 
supported in $[0,R]$. We recall from section
2 the following consequence of the matching condition
\begin{equation}\label{outsideR}
e^{-\lambda(r)}=e^{\mu(r)}=\sqrt{1-\frac{2M}{r}+\frac{Q^2(r)}{r^2}},\;r\geq
R,
\end{equation}
so that $e^{\mu+\lambda}=1$ for $r>R.$ 
Let us now derive an explicit expression for $\mu$.
The Einstein equation (\ref{ee2}) can be written as
\begin{equation}\label{mur}
\mu_r=\big(\frac{m_i}{r^2}+4\pi rp
-\frac{q^2}{2r^3}+\frac{F}{2r^2}\big)e^{2\lambda},
\end{equation}
so that 
\begin{equation}\label{explicitmu}
\mu(r)=-\int_r^\infty\big(\frac{m_i}{r^2}+4\pi rp
-\frac{q^2}{2r^3}+\frac{F}{2r^2}\big)e^{2\lambda}dr,
\end{equation}
since $\mu\to 0$ as $r\to\infty$ in view of (\ref{outsideR}). 
We will also need an explicit formula for $\lambda_r$, and from (\ref{ee1}) we have 
\begin{equation}\label{explicitlambda}
\lambda_r=(4\pi r\rho(r)-\frac{m_i(r)}{r^2}+\frac{q^2}{2r^3}-\frac{F}{2r^2})
e^{2\lambda}. 
\end{equation}
From the expressions of $\mu_r$ and $\lambda_r$ we also obtain 
\begin{equation}
\mu(r)+\lambda(r)=-\int_r^\infty 4\pi \eta(\rho+p)e^{2\lambda}d\eta,
\end{equation}
so that in particular 
\begin{equation}\label{mupluslambda}
\mu+\lambda\leq 0. 
\end{equation}
Now we derive our fundamental integral equation which is a 
consequence of the Tolman-Oppenheimer-Volkov equation. 
Let $$\psi=(m_g+4\pi r^3p-\frac{q^2}{r})e^{\mu+\lambda}.$$
Taking the derivative of $\psi$ with respect to $r,$ a
straightforward calculation using the Tolman-Oppenheimer-Volkov
equation (\ref{TOV}) results in the following equation
\begin{equation}
(m_g+4\pi r^3p-\frac{q^2}{r})e^{\mu+\lambda}= \int_0^r
e^{\mu+\lambda} (4\pi\eta^2
(\rho+p+2p_T)+\frac{q^2}{\eta^2})d\eta.\label{fundamentaleq}
\end{equation}
This equation must be satisfied by any spherically symmetric static solution 
of the Einstein-Maxwell system. 

Let us now consider our sequence of solutions.
Since $p_k(R)=0,\; (m_g)_k(R)=M_k,\, q_k(R)=Q_k$ and $e^{\mu_k+\lambda_k}(R)=1$ we get in view of
(\ref{fundamentaleq}) for $r=R,$ 
\begin{equation}\label{MkQ2}
M_k-\frac{Q_k^2}{R}=\int_{R_k}^{R} e^{\mu_k+\lambda_k}(4\pi\eta^2 (\rho_k+p_k+2(p_T)_k)
+\frac{q_k^2}{\eta^2}) \,d\eta. 
\end{equation}
Here we also used the fact that the matter is supported in $[R_k,R]$. 
We split the right hand side as follows
\begin{eqnarray}
\displaystyle & &\int_{R_k}^{R} e^{\mu_k+\lambda_k}(4\pi\eta^2 (\rho_k+p_k+2(p_T)_k)
+\frac{q_k^2}{\eta^2}) \,d\eta\nonumber\\
\displaystyle & &=\int_{R_k}^{R} e^{\mu_k+\lambda_k}(8\pi\eta^2 \rho_k
+\frac{q_k^2}{\eta^2}) \,d\eta\nonumber\\
\displaystyle & &\phantom{5}+\int_{R_k}^{R} e^{\mu_k+\lambda_k}(4\pi\eta^2 (p_k+2(p_T)_k-
\rho_k) \,d\eta =:S_k+T_k. 
\end{eqnarray}
By the mean value theorem we get that there is a $\xi\in[R_k,R]$ such that 
\begin{eqnarray}\label{kequality}
&\displaystyle S_k= 2e^{\mu_k(\xi)}\xi\int_{R_k}^{R}e^{\lambda_k}(4\pi\eta\rho_k+
\frac{q_k^2}{2\eta^3}) d\eta &\\
&\displaystyle = 2e^{\mu_k(\xi)}\xi\int_{R_k}^{R}[-\frac{d(e^{-\lambda_k})}{d\eta}]d\eta
+2e^{\mu_k(\xi)}\xi\int_{R_k}^{R}(\frac{(m_i)_k(\eta)}{\eta^2}
+\frac{F_k(\eta)}{2\eta^2})e^{\lambda_k} d\eta \nonumber\\
&=:S_k^1 + S_k^2.& 
\end{eqnarray}
Here we used equation (\ref{explicitlambda}) for $\lambda_r$. 
Now, since $\sup_{k}q_k/r$ is strictly less
than one we obtain a uniform bound on $\lambda_k$ from the inequality (\ref{Dan}), 
cf. the consistency check after the formulation of Theorem \ref{thm1}. 
The same computation guarantees that $(m_i)_k(r)/r+F_k(r)/2r<1/2,$ thus it follows that 
$$0\leq S_k^2\leq C\log{\big(\frac{R}{R_k}\big)}\to 0 \mbox{ as }k\to\infty.$$ Since $\mu_k+\lambda_k\leq 0$ by (\ref{mupluslambda}) 
and since $\rho_k\geq p_k+2(p_T)_k \geq 2(p_T)_k,$ it follows from 
the assumptions on the sequence that also 
$$T_k\to 0 \mbox{ as } k\to\infty.$$ 
For the term $S_k^1$ we get by (\ref{outsideR})
\[
S_k^1=-2e^{\mu_k(\xi)}\xi \int_{R_k}^{R}\frac{d}{d\eta}(e^{-\lambda_k})\, d\eta=
2e^{\mu_k(\xi)}\xi\big(1-\sqrt{1-\frac{2M_k}{R}+\frac{Q_k^2}{R^2}}\,\big).
\]
Note here that $\lambda_k(R_k)=0$ due to the support condition of the matter terms. 
In view of (\ref{explicitmu}) and the general bounds on $m_i/r$ and $q/r$ it follows 
that $$e^{\mu_k(\xi)}\to \sqrt{1-\frac{2M}{R}+\frac{Q^2}{R^2}}\mbox{ as }k\to\infty,$$ 
so that
\[
\lim_{k\to\infty}S_k^1=2R\sqrt{1-\frac{2M}{R}+\frac{Q^2}{R^2}}\big(1-\sqrt{1-\frac{2M}{R}+\frac{Q^2}{R^2}}\,\big). 
\]
In conclusion, from (\ref{MkQ2}) we get in the limit $k\to\infty,$ 
\begin{equation}\label{Mineq}
M-\frac{Q^2}{R}=2R\sqrt{1-\frac{2M}{R}+\frac{Q^2}{R^2}}\big(1-\sqrt{1-\frac{2M}{R}+\frac{Q^2}{R^2}}\,\big). 
\end{equation}
After some algebra this relation can be written as
$$M-\frac{Q^2}{R}= (3M-\frac{Q^2}{R})\sqrt{1-2M/R+\frac{Q^2}{R^2}}.$$
Squaring both sides one finds after some rearrangements 
$$(9M^2-\frac{6MQ^2}{R}+\frac{Q^4}{R^2})(\frac{2M}{R}-\frac{Q^2}{R^2})=
4MR(\frac{2M}{R}-\frac{Q^2}{R^2}),$$
so that
\begin{equation}\label{mellanineq}
(3M-\frac{Q^2}{R})^2= 4MR.
\end{equation}
We have thus arrived at the same expression (with equality instead of inequality) 
as (\ref{square}) and we accordingly obtain 
$$\sqrt{M}=\frac{\sqrt{R}}{3}+\sqrt{\frac{R}{9}+\frac{Q^2}{3R}},$$
which completes the proof of Theorem \ref{thmshells}. 
\begin{flushright}
$\Box$
\end{flushright}

\section{Final remarks}
In \cite{KS} several different conditions on the relation between $\rho,\,p$ and $p_T$ 
are investigated, e.g. the isotropic case where $p=p_T.$ We have not tried to 
consider other cases than $p+p_T\leq\rho$ in this work although we believe that it can be done.
We believe however that an equally transparent inequality as (\ref{Dan}) is unlikely to be 
found under other conditions than  $p+p_T\leq\rho,$ cf. \cite{KS}. 
However, the following comparison with the non-charged case is interesting. 
The original Buchdahl inequality \cite{Bu1} was derived 
under the assumptions that $\rho$ is non-increasing outwards and the pressure is isotropic 
and the steady state that saturates the inequality $2M/R\leq 8/9$ within this class 
of solutions is the one with constant energy density for which the pressure is infinite 
at the center. It is quite remarkable that exactly the same inequality holds much more 
generally \cite{An1}, as long as $p+p_T\leq\rho,$ and in particular 
that the steady state that saturates the inequality in this class is an infinitely thin shell 
which is drastically different from the constant energy density solution. One can now ask if 
there is a similar analogue in the charged case. 

In the work \cite{GR} by Giuliani and Rothman they find an explicit solution with constant 
energy density and constant charge density and they obtain for this solution an algebraic 
equation from which the values of the stability radius can be evaluated. 
It is in view of the discussion above 
therefore interesting to see whether these values are less, equal or greater than the values 
given by (\ref{Dan}). 
In \cite{GR} the ratios $R/M$ are displayed for different ratios of $Q/R$ (or more 
precisely for different ratios $Q/M,$ but the corresponding ratios $Q/M$ can be deduced).
It turns out that the critical stability radius given by the relation 
$$\sqrt{M}=\frac{1}{3}+\sqrt{\frac19+\frac{Q^2}{3R}}$$
are smaller than the corresponding ones found in \cite{GR}, or alternatively, 
our relation admits a larger ratio $M/R$ for a given ratio $Q/M.$ 

%A final remark concerns an issue brought up in \cite{KSA} where they discuss the relation 
%between the smallest radius which admits a static solution and the smallest radius for 
%which the optical geometry admits an embedding in the Euclidean space $\mathbb{E}^3$. They  
%notice that in the non-charged case these values coincide, i.e., $R=9M/4$ (assuming 
%the condition $p+p_T\leq\rho$ for the static solution or assuming Buchdahl's original 
%hypothesis on the matter terms), and they therefore raise the question if this holds more 
%generally. That this is not the case has already been pointed out in \cite{dFSY2} using 
%a semi-analytic argument and our result in this work also shows that there is no agreement in 
%the charged case. 

\begin{center}
\textbf{Acknowledgement}
\end{center}
I would like to thank the authors of \cite{GR} for their clearly written paper
which got me interested in this topic.

\end{document}